---------------------------------------------------------------
\documentstyle[12pt]{article}
\begin{document}

\title{\bf Transport and current reversal in stochastically
driven ratchets}

 \author{Mark M. Millonas \\ \small
Complex Systems Group and Center for Nonlinear   Studies, Theoretical
Division,\\ \small MS B258   Los Alamos
National Laboratory, Los Alamos,   NM 87545\\ \small \&  Santa Fe
Institute,
 Santa Fe, NM 87501 \\ \and
Mark I. Dykman \\  \small
Department of Physics, Stanford University, Stanford, CA 94306.
}

\maketitle

\begin{abstract}

We present analytic results for the current in a system moving
in an arbitrary periodic potential and driven by weak Gaussian
noise with an arbitrary power spectrum which are valid  to
order $(t_c/t_r)^2$, where $t_c$ is the largest characteristic
time of the noise, and $t_r$ is the characteristic intrawell
relaxation time.  The dependence of the current on the shape of
the potential, and on the shape of the power spectrum of  the
noise  is illustrated. It is demonstrated that the direction of
the current is opposite when the power spectrum of the noise
has minimum or maximum at zero frequency. A simple physical
mechanism for this behavior is suggested.  The behavior of the system
in the limit of slow noise ($t_c >>t_r$) is also discussed.

\end{abstract}

Recently some intriguing work has  appeared on the subject of stochastically
driven
ratchets,\cite{Man,Doering}
and the transformation of noise in spatially-periodic (or
phase-periodic) systems into a current.
Such systems are noise rectifiers
and thus  are of great interest in
relation to the understanding of machinery which operates in
\lq\lq the Brownian regime" \cite{note} that is, on the very
small scale where fluctuations play a major role, and where the
basic macroscopic methods of
controlling energy flow no longer remain valid.   The most
immediate examples are biological systems,\cite{bio} but these
ideas may not be irrelevant in the area of applied technology,
where a great interest has developed recently in the possible
construction of nanoscale devices.\cite{nano}

A simple model for a ratchet is a noise-driven
overdamped nonlinear dynamical system described by the
stochastic differential equation
\begin{equation} \dot x = - U^\prime(x) +
f(t), \ \ \ \quad U(x) = U(x+\lambda),\ \ \  \langle f(t)\rangle = 0,
\end{equation} where $U(x)$ is a periodic potential, such as the one
illustrated in Fig. 1, and $f(t)$ is zero-mean noise of some type.  The
theoretical problem is to find the stationary current density $j = \langle
\dot{x}(t)\rangle$ in the ratchet given the shape of $U(x)$ and the
properties of the noise $f(t)$, and then to find the most appropriate
conditions for the transformation of the noise into the current. If
the noise is white then the system is (quasi)thermal, the second law
of thermodynamics applies, and $j=0$. If the noise is not white, i.e.,
for colored noise, the system is no longer in thermal equilibrium, and
in general $j \neq 0$. Since onset of a current means breaking the
\lq\lq right-left" symmetry, currents may only arise, in the case of
additive noise, if the potential $U(x)$ is asymmetric with respect to
its extrema. The onset of a current can be viewed also as an example
of \lq\lq temporal order coming out of disorder", since the current is
apparently time-irreversible, whereas stationary noise does not
distinguish \lq\lq future" from the \lq\lq past"; we notice, however,
that Eq. (1) implies relaxation and is thus time-irreversible itself.

We consider the case  where $f(t)$ is zero-mean Gaussian noise
with a frequency-dependent power spectrum
\begin{equation}
\Phi(\omega) = \int^\infty_{-\infty} dt\ \exp(i\omega t)\phi(t),\ \ \
\phi(t)=<f(t)f(0)>.
\end{equation}
If the characteristic noise intensity $D = {\rm max}\,\Phi
(\omega)$ is small then most of the time the system
performs small-amplitude fluctuations about the minima of the
potential. Occasionally it \lq\lq jumps" from the minimum it
occupied to the  one on the right or left, with the
probabilities per unit time $W_+$ and $W_-$, respectively.
These jumps give rise to the current $j = \lambda (W_+ - W_-)$.
The dependence of the transition probabilities $W_{\pm}$ on the
noise intensity $D$ is of the activation type for Gaussian
noise, $W_{\pm} = {\rm const.}\times \exp(-R_{\pm}/D)$. The
characteristic activation energies $R_{\pm}$ of the escapes
from a given minimum $x_0$ of the potential over the right
($x_+$) or left ($x_-$) maximum are given by the solution of
a variational problem formulated in \cite{mark}. In the case
where the bandwidth of the spectrum $\Phi (\omega)$ greatly
exceeds the reciprocal relaxation time of the system $t_r^{-1}
= U^{\prime\prime}(x_0)$ the activation energies are $R_{\pm} = 2
F(0)\Delta U+ \gamma_{\pm}\ F^{\prime\prime}(0)$, where
\begin{equation} \Delta U =U(b) - U(a)>0 ,\ \ \ \
\gamma_{\pm}=\int_{x_0}^{x_{\pm}}  dx\
U^\prime(x)[U^{\prime\prime}(x)]^2\geq 0,
\end{equation}
\begin{equation} F(\omega) =D/\Phi(\omega),\ \ \
F^{\prime\prime}(\omega) = {d^2F(\omega)\over d\omega^2},\ \ \
\vert F^{\prime\prime}(0)/F(0)\vert<<t^2_r.
\end{equation}
The current is then given by
\begin{equation}
j =
 \lambda W_K [\exp(-
\gamma_+F^{\prime\prime}(0)/D) -\exp(-\gamma_-
F^{\prime\prime}(0)/D)],
\end{equation}
where $W_K$ is the Kramer's activation rate
\begin{equation}
W_K={{\pi D}/ F(0)\over\sqrt{
U^{\prime\prime}(x_0) |U^{\prime\prime}(x_+)|}}\exp(-2 F(0)\Delta
U/D).
\end{equation}
 Eq. 5 is a principle result of this letter. It is immediately obvious
from (5) that: (i) the noise color does give rise to the onset of a
current due to the fluctuational interwell transitions, and (ii) the
direction of the current depends crucially on {\it the shape of the
spectral density} since $F^{\prime\prime}(0)$ can take both positive
or negative signs. We emphasize that, although the corrections
$\gamma_{\pm}F^{\prime\prime}(0)$ to $R_{\pm}$ are small compared to
the main term, they are not small compared to the noise intensity $D$
and can change $W_{\pm}$ by orders of magnitude (we have neglected the
corrections to the prefactor in $W_{\pm}$ due to the noise color and
used the standard Kramers expression for this prefactor valid for
white-noise driven systems). In fact, except the special case where
$U(x)$ is symmetric with respect to $x_0$, the ratio of the probabilities
$W_+/W_-$ is {\it exponentially} large or small for small intensity of
the colored noise, and therefore the transitions in one direction
dominate overwhelmingly over the transitions in the opposite
direction, so that $|j| \approx \lambda W_>$ where $W_> = {\rm
max}(W_+, W_-)$. The direction of the current is determined by the
interplay of the shape of the potential and the features (the shape of
the power spectrum, in the present case) of the noise.

The dependence of the current on the noise color is different
depending on the sign of $F^{\prime\prime}(0)$. When
$F^{\prime\prime}(0) >0$ there is a saturation effect where, given the
noise strength $D$ and the shape of the potential, the current is
maximal for $F^{\prime\prime}(0)$ given by

\begin{equation}
F^{\prime\prime}_m(0) = {D\ln(\gamma_- /\gamma_+)\over
(\gamma_--\gamma_+)},\
\ \  F^{\prime\prime}_m(0)>0.\end{equation}
Thus, for $F^{\prime\prime}(0)>0$,  $F^{\prime\prime}_m(0)$ is the optimal
noise color for a given noise strength and ratchet potential.

Eq.(5) also provides an answer to the following question: given the
barrier height $\Delta U$, what is the shape of the well for which the
current (5) will be most pronounced for weak noise color, that is,
what is the most effective shape of the ratchet? This shape is given by
the extreme value of $\gamma_{\pm}$ for given $\Delta U$ and period
$\lambda$.  It is straightforward to show that the corresponding
variational problem does not have a differentiable solution: the
single-valued potential has to be of the shape approaching the
one pictured in Fig. 2.  A simple
analysis can be done, e.g., for $U^{\prime}(x) = A_1 + A_2
\tanh [(x - \tilde{x}_0)/a], \; -\lambda/2 < x < \lambda/2$
($A_1 + A_2
\tanh [(x_0 - \tilde{x}_0)/a] = 0$):
for small $a/\lambda$ the shape of the corresponding potential
is close to that of the sawtooth.

The above results apply immediately to the most often
studied  case where $f(t)$ is exponentially correlated,
\begin{equation}
\phi(t) = {D\over 2\tau}\exp(-\vert t \vert/\tau),\ \ \ \
\Phi(\omega) =  D/(1+\omega^2\tau^2).
\end{equation}
The problem of fluctuational transitions induced by this noise has
been investigated for small noise intensities in very much detail (see
\cite{McKane}, \cite{Tsironis}, \cite{Bray} and also \cite{DL} for a review).
The noise (8) has one correlation time, $\tau$, and
$F^{\prime\prime}(0) = 2\tau^2 > 0$.  Obviously, $j$ vanishes to first
order in $\tau$, in agreement with the results of [2]. It has a
maximum as a function of $\tau$ for $\tau$ given by (7), and falls
down for large $\gamma_{\pm}\tau^2/D$.  The saturation of $j$ vs.
$\tau$ is illustrated in Fig. 3.

Naturally occurring noise will generally not be exponentially
correlated, a situation which any realistic physical theory must
accommodate. This is clearly the regime in which $j\neq 0$. The
advantage of the result presented here is that it not only gives an
analytic result to second order in $t_c/t_r$, but is valid for any
Gaussian noise. A more general situation  than (8)
can be modeled \cite{DL} by the noise with the power
spectrum
\begin{equation}
\Phi(\omega) = {4\Gamma \tilde{D}\over (\omega^2 -\omega^2_0)^2+ 4
\Gamma^2\omega^2},
\end{equation}
 where $\Gamma$ is a measure of the bandwidth, and
$\omega_0$ the frequency of the noise. This  can be thought of
as the power spectrum of the system,
\[\ddot f + 2 \Gamma \dot f + \omega_0^2 f = \xi(t) \]
\begin{equation}<\xi(t)\xi(s)>= 4\Gamma \tilde{D}\delta(t-s).
\end{equation}
 In this case $F(0)= \omega^4_0 D/4\Gamma\tilde{D}$,
and $F^{\prime\prime}(0)= (2\Gamma^2 - \omega^2_0)D/\Gamma
\tilde{D}$. Obviously, $F^{\prime\prime}(0) > 0$ when
$\omega_0^2 < 2\Gamma^2$ and thus the maximum of the power
spectrum (8) is at $\omega = 0$. In this case the direction of
the current for $\Gamma \gg t_r^{-1}$ is the same as for
exponentially correlated noise and the analysis given above
directly applies.

A completely different situation occurs if $\omega_0^2 > 2\Gamma^2$
and the power spectrum (8) has a {\it minimum} at $\omega = 0$. In
this case $F^{\prime\prime}(0) < 0$. It follows from (5) that the
direction of the current is {\it opposite} to that arising for
exponentially correlated noise.  In contrast to the case
$F^{\prime\prime}(0) > 0$ considered above, in the present case
increase in $|F^{\prime\prime}(0)|$ does not give rise to saturation
and then to decrease in $j$: the current is increasing exponentially
with the increasing $|F^{\prime\prime}(0)|$ where the approximation
(3) is applicable. The current reversals have been found recently in
numerical experiments, and in certain  specific exactly solvable
cases, by Doering and Horsthemke\cite{Doering}. No plausible physical
mechanism has been suggested as an explanation in \cite{Doering}. Not
only does the present analysis give an analytic criterion for the
current reversal, $\Phi^{\prime\prime}(0) \propto -
F^{\prime\prime}(0) > 0$, but it also suggests a direct physical
interpretation.

The correction to the activation energy of a transition $R$ in the
case of \lq\lq weakly colored" noise is due to the fact that it is not
only the total work the noise does on the system on the way from the
potential minimum to the barrier top (along the optimal path of the
escape) that counts, as in the case of white noise, but the
characteristic strength and duration of the pulse that gives rise to
the escape. The parameters $\gamma_{\pm}$ just characterize the ratio
of the squared value of the force to the duration of the pulse (cf.
\cite{mark}). The shorter the pulse the higher are the
frequencies of the noise components involved. If
the power  spectrum of the noise $\Phi(\omega)$ decreases with $\omega$
($F^{\prime\prime}(0) > 0$) then the higher-frequency
noise components are weaker on the average, and the
escape probability decreases
with the decreasing duration of the pulse.  For
$\Phi(\omega)$ increasing with $\omega$ ($F^{\prime\prime}(0) < 0$) the
result
is exactly  opposite. The characteristic duration of the pulse is
determined by the ratio of the distance between the potential
minimum and the barrier top to the characteristic velocity
$U^{\prime}$, and for a given height of the potential barrier
it scales as $( U^{\prime})^{-2}$.

Suppose we have a
potential more steep to the left from the minimum, as shown in
Fig.1. It is clear from (3) that in this case $\gamma_- >
\gamma_+$. The pulse which gives rise to the escape over the
left barrier is shorter, and the characteristic frequencies
involved are higher. Therefore, for $F^{\prime\prime}(0) > 0$
(as in the case of exponentially correlated noise), the escape
over the left barrier is less likely to happen than that over
the right one, and the current flows to the right, whereas for
$F^{\prime\prime}(0) < 0$ it flows to the left.

The behavior of the system when driven by slow noise ($t_c >> t_r$) is
also of interest.  The activation energies in this limit can also be
calculated via the method of Ref. \cite{mark}, and are given by $R_\pm
= D [U^\prime_{m,\pm}]^2/2\phi (0)$ where $\vert
U^\prime_{m,\pm}\vert$ are the maximum values of $\vert U^\prime\vert$
on the intervals $(x_0,x_\pm)$.  We note here in passing that there
are nonanalytic [$(t_r/t_c)^{2/3}]$ corrections to $R_\pm$ \cite{mark,Bray}.
The current which arises in the case of slow noise is then
\begin{equation}
j = \lambda [C_+\exp(-
[U^\prime_{m,+}]^2/2\phi (0))- C_-\exp(-
[U^\prime_{m,-}]^2/2\phi (0))].
\end{equation}
($C_{\pm}$ are the constants that allow for the prefactors in the
expressions for $W_{\pm}$; obviously, $\phi (0)$ is just the
mean-square value of the noise). What is happening here also has a
clear physical interpretation (cf. \cite{Tsironis}).  Since the noise
has an extremely long correlation time, in the range where
$U^{\prime\prime}(x) > 0$ the particle simply follows the force
adiabatically according to $U^\prime(x(t)) = f(t)$.  The fluctuation
large enough to allow the particle to escape over a barrier is just
the one that overcomes the restoring force $- U^{\prime}(x)$ for all
$x$, and the probability of such a fluctuation is just $W\propto
\exp(-R/D)$.  The current in this case is always in the positive
direction for $[U^\prime_{m,-}]^2>[U^\prime_{m,+}]^2$.  There is again
a saturation effect, and the current is maximized for
$ \phi^{-1}(0) \approx
{ 4 D \ln(\vert C_- U^\prime_{m,-}/ C_+ U^\prime_{m,+}\vert)/
([U^\prime_{m,-}]^2-[U^\prime_{m,+}]^2)}.$

We emphasize that the onset of current in the system considered is the
result of it being away from thermal equilibrium. This means that
dissipation and fluctuations are not interrelated via
fluctuation-dissipation theorem. In the particular case considered
dissipation was not retarded (the friction force is determined by the
instantaneous value of the coordinate), and therefore the effect
arises when the power spectrum of the noise displays dispersion (for
Gaussian noise).  A current can also arise  in the situation
of a white-noise driven system where dissipation is retarded.

Lastly we remark that the system can be expected to exhibit behavior similar to
stochastic resonance as the noise strength $D$ is varied, with $j$ exhibiting a
maximum for some noise strength.  Obviously for very small $D$ the current
is an increasing function of $D$.  As the noise strength is increased the
fluctuations will begin to wash out the effects due to the shape of the
potential, leading to a decrease of the current for large $D$.

\begin{figure}[t]
\vspace{.5in}
\hspace{0in}\special{postscriptfile pot scaled 1400}
\caption{\sf Typical ratchet potential $U(x)$.}
\end{figure}

\begin{figure}[t]
\vspace{.5in}
\hspace{1in}\special{postscriptfile opt scaled 1100}
\caption{\sf The  sawtooth potential which is the ``optimal" limit of $U(x)$
which gives
rise to the greatest current.}  \end{figure}

\begin{figure}[t]
\vspace{.5in}
\hspace{1in}\special{postscriptfile sat scaled 1000}
\caption{\sf Current vs. $\tau$ for varying potential shape showing that the
current is maximized for a specific $\tau$.}  \end{figure}

\end{document}